\documentclass[epj,10pt]{svjour}

\usepackage{tikz}

\usepackage{graphics}
\usepackage{hyperref}
\hypersetup{colorlinks = true, allcolors = blue}
\usepackage{authblk}
\usepackage{hepparticles}
\usepackage{hepunits}
\usepackage{hepnames}
\usepackage{amsmath}
\usepackage{amssymb}
\usepackage{multicol}
\usepackage{multirow}
\usepackage{xspace}
\usepackage{color, colortbl}
\definecolor{Gray}{gray}{0.9}
\definecolor{LightCyan}{rgb}{0.88,1,1}

\usepackage{enumitem}
\setlist[description]{leftmargin=\parindent,labelindent=0.25cm}

\newcommand{\epem}{\mathrm{e^+e^-}}
\newcommand{\gaga}{\gamma\,\gamma}
\newcommand{\sqrts}{\sqrt{s}}
\newcommand{\LumiInt}{\mathcal{L}_\mathrm{\tiny{int}}}
\newcommand{\bbbar}    {b\overline{b}}
\newcommand{\ccbar}    {c\overline{c}}
\newcommand{\qqbar}    {q\overline{q}}

\newcommand{\pt}{p_{_{\mathrm{T}}}}
\newcommand{\mT}{m_{_{\mathrm{T}}}}

\newcommand*{\taulep}{\ensuremath{\tau_\mathrm{lep}}\xspace}
\newcommand*{\tauhad}{\ensuremath{\tau_\mathrm{had}}\xspace} 
\mathchardef\mhyphen="2D

\providecommand{\MG}{{\sc MadGraph\,5}}

\providecommand{\pythia}{{\sc pythia}}
\providecommand{\whizard}{{\sc whizard}}
\providecommand{\kkmc}{{\sc kkmc}}
\newcommand{\hdecay}{{\sc hdecay}}
\newcommand{\fastjet}{{\sc FastJet}}

\newcommand{\com}{c.m.\@\xspace}

\newcommand{\ie}{i.e.\@\xspace}
\newcommand{\eg}{e.g.\@\xspace}

\DeclareMathSizes{10}{10}{6}{4}

\begin{document}

\title{Measuring the electron Yukawa coupling via resonant $s$-channel Higgs production at FCC-ee}
\author{David d'Enterria\inst{1}\thanks{Corresponding author: david.d'enterria@cern.ch} 
\and Andres Poldaru\inst{2} \and George Wojcik\inst{3}
}                     
\institute{CERN, EP Department, 1211 Geneva, Switzerland \and LMU, 80539 Munich, Germany \and SLAC, 2575 Sand Hill Rd., Menlo Park, CA, 94025 USA}
%
%
\date{Received: \today / Revised version: \today }
%
\abstract{
The Future Circular Collider (FCC-ee) offers the unique opportunity of studying the Higgs Yukawa coupling to the electron, $y_\mathrm{e}$, via resonant $s$-channel production, $\epem \to \mathrm{H}$, in a dedicated run at $\sqrt{s} = m_\mathrm{H}$. The signature for direct Higgs production is a small rise in the cross sections for particular final states, consistent with Higgs decays, over the expectations for their occurrence due to Standard Model (SM) background processes involving $\mathrm{Z}^*$, $\gamma^*$, or $t$-channel exchanges alone. Performing such a measurement is remarkably challenging for four main reasons. First, the low value of the e$^\pm$ mass leads to a tiny $y_\mathrm{e}$ coupling, and correspondingly small cross section: $\sigma_\mathrm{ee\to H} \propto m_\mathrm{e}^2 = 0.57$~fb accounting for initial-state $\gamma$ radiation. Second, the $\epem$ beams must be monochromatized such that the spread of their center-of-mass (\com) energy is commensurate with the narrow width of the SM Higgs boson, $\Gamma_\mathrm{H} = 4.1$\,MeV, while keeping large beam luminosities. Third, the Higgs mass must also be known beforehand with a few-MeV accuracy in order to operate the collider at the resonance peak, $\sqrt{s} = m_\mathrm{H}$. Last but not least, the cross sections of the background processes are many orders-of-magnitude larger than those of the Higgs decay signals. A preliminary generator-level study of 11 Higgs decay channels using a multivariate analysis, which exploits boosted decision trees to discriminate signal and background events, identifies two final states as the most promising ones in terms of statistical significance: $\mathrm{H}\to gg$ and $\mathrm{H}\to \mathrm{W}\mathrm{W}^*\!\to \ell\nu$\,+\,2\,jets. For a benchmark monochromatization with 4.1-MeV \com\ energy spread (leading to $\sigma_\mathrm{ee\to H} = 0.28$~fb) and 10\,ab$^{-1}$ of integrated luminosity, a $1.3\sigma$ signal significance can be reached, corresponding to an upper limit on the e$^\pm$ Yukawa coupling at 1.6 times the SM value: $|y_\mathrm{e}|<1.6|y^\mathrm{\textsc{sm}}_\mathrm{e}|$ at 95\% confidence level, per FCC-ee interaction point per year. Directions for future improvements of the study are outlined.
\PACS{
      {14.80.Bn}{Standard Model Higgs boson}   \and
      {14.60.Cd}{electrons and positrons} 
     } 
} 

\authorrunning{D. d'Enterria \textit{et al.}}
\titlerunning{Electron Yukawa coupling via $s$-channel Higgs production at FCC-ee}

\maketitle
%


\section{Introduction}
\label{sec:intro}

The usual claim that the Brout--Englert--Higgs mechanism of mass generation for elementary particles~\cite{Englert:1964et,Higgs:1964ia,Higgs:1964pj} has been experimentally confirmed at the Large Hadron Collider (LHC) thanks to the Higgs boson discovery~\cite{Aad:2012tfa,Chatrchyan:2012ufa}, and subsequent studies of its properties~\cite{Gray:2020sam,Bass:2021acr}, is only valid so far for the heaviest Standard Model (SM) particles: W and Z weak bosons, and quarks and leptons of the third family ($t$, $b$, and $\tau$). Today, not only the generation of all neutrino masses remains a mystery~\cite{Deppisch:2015qwa}, but  at the end of the LHC lifetime only a fraction of the Higgs Yukawa couplings to the second-family fermions (the muon and, maybe, the charm quark) will have been probed. On the other hand, due to their low masses and thereby small Yukawa couplings to the Higgs field, the mass generation mechanism for the stable matter of the visible universe, composed of $u$ and $d$ quarks 
plus the electron and neutrinos ($\nu$), will remain experimentally untested. The smallest Yukawa coupling, aside from the Dirac $\nu$'s case, is that of the electron given by $y_\mathrm{e} = \sqrt{2} m_\mathrm{e}/v = 2.9\cdot10^{-6}$ for $m_\mathrm{e}= 0.511\cdot10^{-3}$\,GeV 
and Higgs vacuum expectation value $v = (\sqrt{2}\mathrm{G_F})^{-1/2} = 246.22$\,GeV. Measuring the Higgs coupling to the electron is impossible at hadron colliders because the $\mathrm{H} \to \epem$ decay has a tiny branching fraction of $\mathcal{B}(\mathrm{H}\to\epem) = 
5.22\cdot10^{-9}$ (see Eq.~(\ref{eq:Gamma_H_ee}) below), and 
is completely swamped by a Drell--Yan $\epem$ continuum whose cross section is many orders of magnitude larger. Measurements in p-p collisions at the LHC, assuming the SM Higgs production cross section, lead to an upper bound on the branching fraction of $\mathcal{B}(\mathrm{H}\to\epem) < 3.6\cdot 10^{-4}$ at 95\% confidence level (CL), corresponding to an upper limit on the Yukawa coupling $y_\mathrm{e}\propto \mathcal{B}(\mathrm{H}\to\epem)^{1/2}$ of 260 times the SM value~\cite{Khachatryan:2014aep,ATLAS:2019old}. Such a constraint can be translated into a lower bound on the energy scale of any physics beyond the SM (BSM) affecting $y_\mathrm{e}$, of 
 $\Lambda_{\textsc{bsm}} \approx 
v^{3/2}(\sqrt{2}m_\mathrm{e}\cdot(y_\mathrm{e}/y^\mathrm{\textsc{sm}}_\mathrm{e}))^{-1/2}\gtrsim 8.8$~TeV~\cite{Altmannshofer:2015qra}. Assuming that the sensitivity to the $\mathrm{H} \to \epem$ decay scales with the square root of the integrated luminosity, the high-luminosity LHC phase with a $\LumiInt = 3\,\mathrm{ab}^{-1}$ data sample~\cite{Cepeda:2019klc} will result in $y_\mathrm{e}\lesssim 120 y^\mathrm{\textsc{sm}}_\mathrm{e}$ (\ie\ $\Lambda_{\textsc{bsm}}\gtrsim 13$~TeV).

The possibility of studying resonant Higgs production at leptons colliders has been considered in the literature so far only for $\mu^+\mu^-$ annihilation at $\sqrt{s} = m_\mathrm{H}$, notably as a means to directly and precisely measure $\Gamma_\mathrm{H}$, $m_\mathrm{H}$, and the muon Yukawa coupling, by exploiting a large peak production cross section of $\sigma_{\mu\mu\to\mathrm{H}} = 70$~pb~\cite{Barger:1996jm}. The same measurement at an $\epem$ machine had never been seriously considered given the sub-femtobarn cross section for the $\epem\to\mathrm{H}$ process, suppressed by at least a factor 
$m_\mathrm{e}^2/m_\mu^2$ compared to the muon collider case. 
Notwithstanding this difficulty, when the FCC-ee was first proposed~\cite{Gomez-Ceballos:2013zzn}, it was noticed that the unparalleled integrated luminosities of about $\LumiInt = 10$\,ab$^{-1}$/year available at $\sqrts = 125$\,GeV, would make it possible to attempt an observation of the direct production of the scalar boson~\cite{dEnterria:2014,dEnterria:2017dac}. Such a consideration motivated a few subsequent works on various $\epem\to\mathrm{H}$ theoretical~\cite{Altmannshofer:2015qra,Jadach:2015cwa,Greco:2016izi,Dery:2017axi} and accelerator~\cite{Zimmermann:2017tjv,ValdiviaGarcia:2019ezi} aspects. 

\begin{figure}[htpb!]
\centering
\includegraphics[width=0.55\columnwidth]{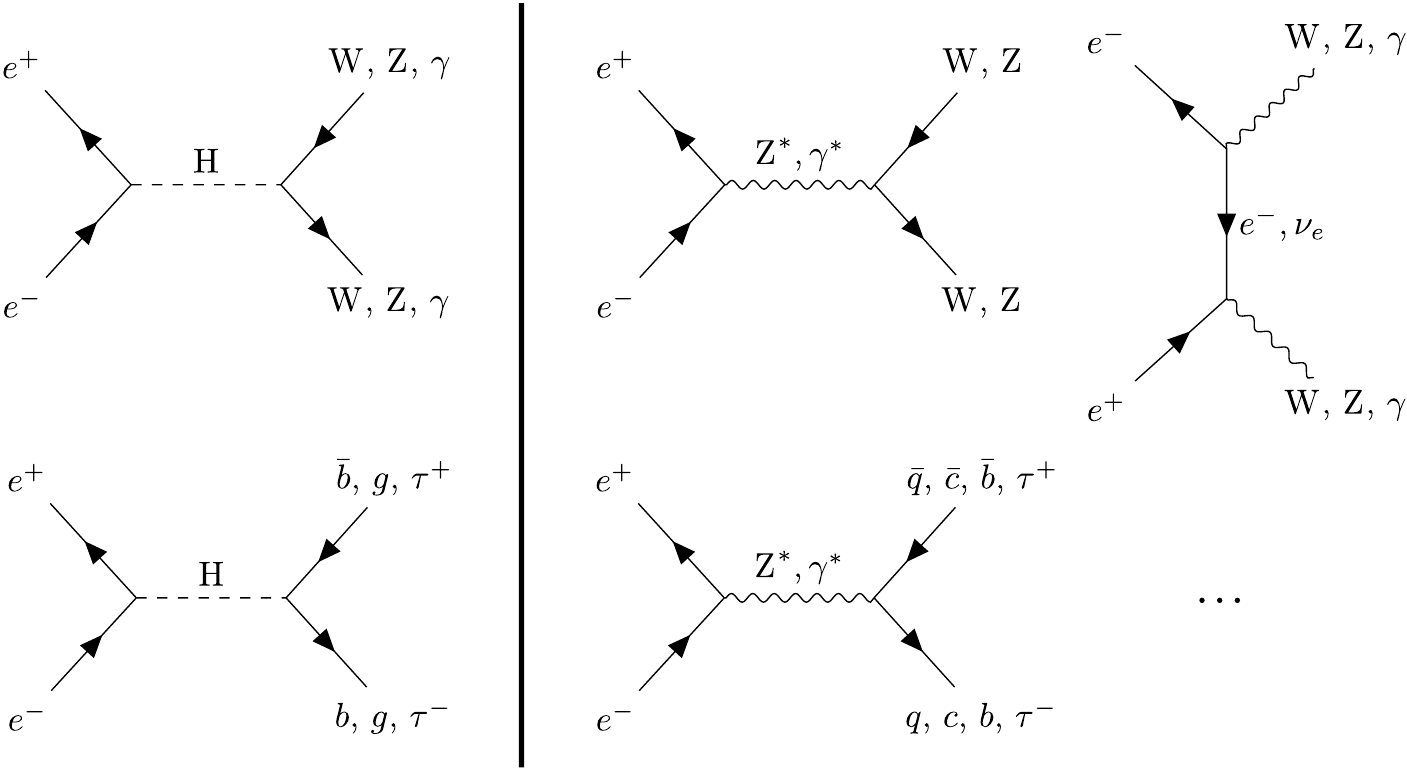}\hspace{0.1cm}
\includegraphics[width=0.425\columnwidth,height=5.5cm]{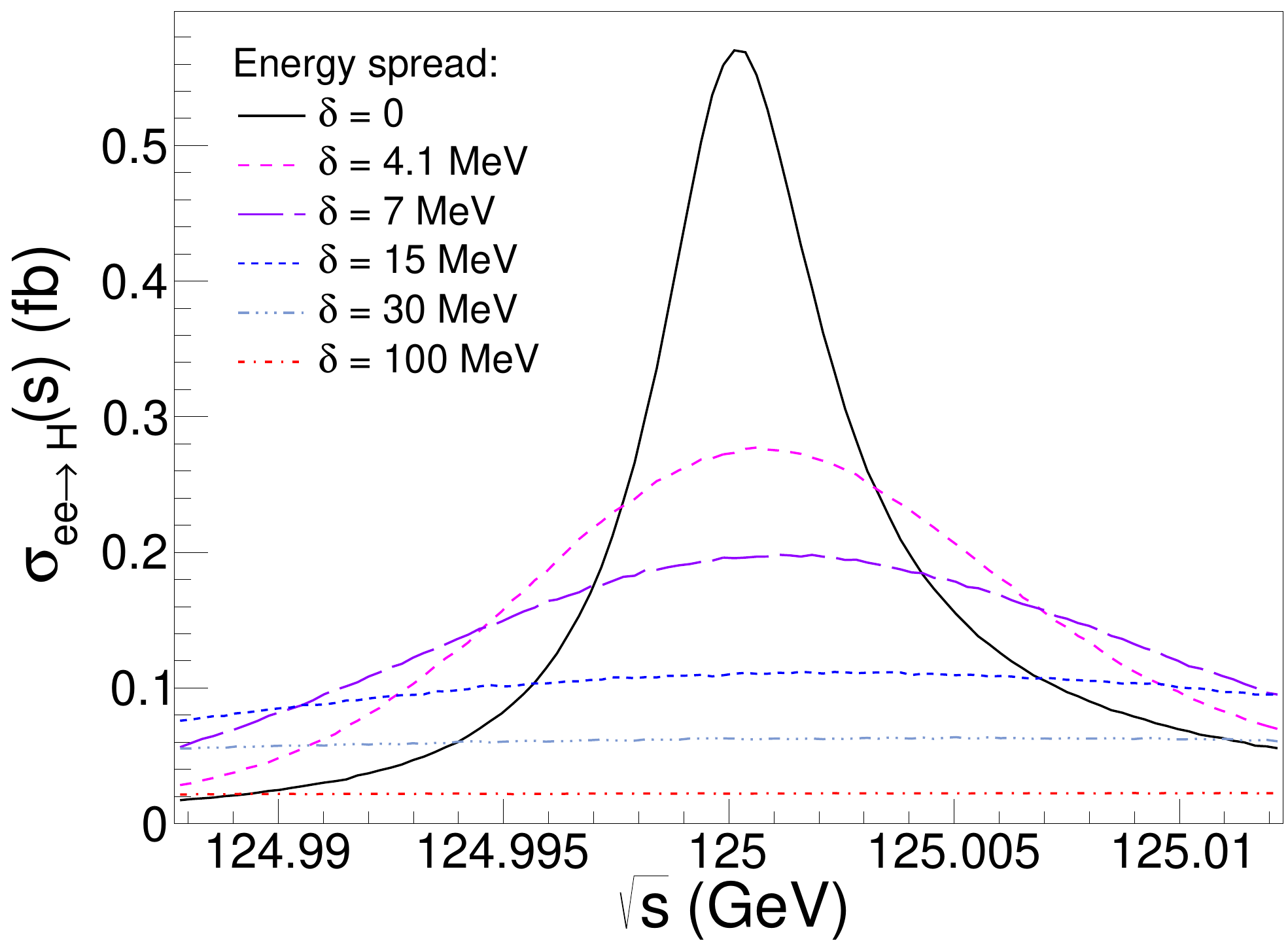}
\caption{Typical diagrams for the direct Higgs channel production (left) 
decaying into electroweak bosons (top) and fermions or gluons (bottom), and associated backgrounds (center), considered in this work. Right: Resonant Higgs production cross section, including ISR effects, for several values of the $\epem$ \com\ energy spread $\delta_{\sqrts}$ = 0, 4.1, 7, 15, 30, and 100\,MeV~\cite{Jadach:2015cwa}.
\label{fig:diags}}
\end{figure}

The Feynman diagram for $s$-channel Higgs production (and dominant decays) is shown in Fig.~\ref{fig:diags} (left). Other $\epem\to\mathrm{H}$ production processes, through W and Z loops, are suppressed by the electron mass for on-shell external fermions and have negligible cross sections~\cite{Altmannshofer:2015qra}. 
The resonant Higgs cross section at any given \com\ energy $\sqrts$ is theoretically given by the relativistic Breit--Wigner (BW) expression:
\begin{equation}
\sigma_\mathrm{ee\to H} = \frac{4\pi\Gamma_\mathrm{H}\Gamma(\mathrm{H}\to\epem)}{(s-m_\mathrm{H}^2)^2 + m_\mathrm{H}^2\Gamma_\mathrm{H}^2},
\label{eq:sigma_H_ee}
\end{equation}
where $\Gamma_\mathrm{H} = 4.1$\,MeV is the total Higgs width~\cite{deFlorian:2016spz}, $m_\mathrm{H}=125$\,GeV its mass, and the partial decay width  $\Gamma(\mathrm{H}\to\epem)$, given by the tree-level relation
\begin{equation}
\Gamma(\mathrm{H}\to\epem) = 
\frac{\mathrm{G_F}m_\mathrm{H}m_\mathrm{e}^2}{4\sqrt{2}\,\pi}\left(1-\frac{4\,m^2_\mathrm{e}}{m^2_\mathrm{H}}\right)^{3/2} = 2.14\cdot10^{-11}\,\mathrm{GeV\,,}
\label{eq:Gamma_H_ee}
\end{equation}
is tiny due to its dependence on the square of the $\mathrm{e}^\pm$ mass. 
From the BW expression (\ref{eq:sigma_H_ee}), it is clear that an accurate knowledge of the $m_\mathrm{H}$ value is critical to maximize the resonant cross section.
Combining three $\epem\to \mathrm{H}\mathrm{Z}$ measurements at FCC-ee (recoil mass, 
peak cross section, and threshold scan), a $\mathcal{O}$(2\,MeV) mass precision is  achievable~\cite{HiggsMassFCCee} before a dedicated $\epem\to\mathrm{H}$ run. In addition,  the FCC-ee beam energies will be monitored with a relative precision of $10^{-6}$~\cite{Blondel:2019jmp}, warranting a sub-MeV accuracy of the exact point in the Higgs lineshape being probed at any moment.
Taking $m_\mathrm{H} = 125$\,GeV, Eq.~(\ref{eq:sigma_H_ee}) gives $\sigma_\mathrm{ee\to H} = 4\pi\mathcal{B}(\mathrm{H}\to\epem)/m_\mathrm{H}^2 = 1.64$~fb as peak cross section. Two effects, however, lead to a significant broadening of the Born-level result: (i) initial-state $\gamma$ radiation (ISR) reduces the cross section and generates an asymmetry of the Higgs lineshape, and (ii) the actual beams are never perfectly monoenergetic, \ie\ the collision $\sqrts$ has a spread $\delta_{\sqrts}$ around its center value, 
further leading to a smearing of the BW peak. The reduction of the BW cross section due to IS photon emission(s) is of factor of 0.35 and leads to $\sigma_\mathrm{ee\to H} = 0.57$~fb~\cite{Jadach:2015cwa}. The additional impact of a given \com\ energy spread on the Higgs BW shape can be quantified through the convolution of BW and Gaussian distributions, \ie\ a relativistic Voigtian function~\cite{Kycia:2018hyf}. Figure~\ref{fig:diags} (right) shows the Higgs lineshape for various $\delta_{\sqrts}$ values. The combination of ISR plus $\delta_{\sqrts} = \Gamma_\mathrm{H} = 4.1$\,MeV reduces the peak Higgs cross section by a total factor of 0.17, down to $\sigma_\mathrm{ee\to H} = 0.28$~fb.
 As a baseline study, we will use this latter value as our default expectation for the signal production cross section, and compute the corresponding significance for a 1-year operation with 10\,ab$^{-1}$ integrated luminosities per FCC-ee interaction point (IP). The computed signal yields and associated significances can then be subsequently rescaled to any other choice of 
$(\delta_{\sqrts}, \LumiInt)$ values given by the chosen beam monochromatization scheme~\cite{Zimmermann:2017tjv,ValdiviaGarcia:2019ezi}. 


\section{Analysis strategy. Simulation of signal and background processes}
\label{sec:xsections}

The strategy to observe the resonant production of the Higgs boson is based on identifying final states in $\epem$ collisions at $\sqrts = m_\mathrm{H}$, consistent with any of the H decay modes, that lead to a small increase (but, hopefully, statistically significant when combined together) of the measured cross sections with respect to the theoretical expectation for their occurrence 
via background processes alone, involving $\mathrm{Z}^*$, $\gamma^*$, or $t$-channel exchanges (Fig.~\ref{fig:diags}, center diagrams). The assumption is that, after various years of FCC-ee operation at the Z pole and HZ \com\ energies~\cite{Abada:2019lih,Abada:2019zxq}, the theoretical knowledge of the overwhelming background cross sections will be at the $10^{-5}$ level or better~\cite{Blondel:2019vdq}, and that experimental systematic uncertainties (detector acceptance, reconstruction efficiencies, luminosity, etc.) will be controlled at the same level of precision~\cite{Abada:2019zxq,Jadach:2018jjo} and/or will partially cancel out in ratios of number of signal over backgrounds yields. Under such circumstances, the proposed measurement can be considered as a very-rare ``counting experiment'' that aims at adding up the individual statistical significances for various final states consistent with known Higgs decay channels in the hope to observe an excess above the background counts expectations.

In order to carry out our simulation studies, we generate individual samples of 10$^5$--10$^7$ $\epem$ annihilation events at $\sqrts = m_Z = 125.00$\,GeV with the \pythia~8 Monte Carlo (MC) code~\cite{Sjostrand:2007gs}, for each of the 11 final states for signal and associated backgrounds listed in Table~\ref{tab:Sigma_Higgs}. The Higgs decay branching fractions used are those from the \hdecay\ code at NLO accuracy~\cite{Djouadi:2018xqq}. The \pythia~8 signal cross sections are absolutely normalized to match our benchmark $\sigma_\mathrm{ee\to H} = 0.28$~fb value for ISR plus $\delta_{\sqrts} = 4.1$-MeV energy spread discussed above (second curve of Fig.~\ref{fig:diags} right). Higgs decay modes not listed in Table~\ref{tab:Sigma_Higgs} are either completely swamped by background (\eg\ $\mathrm{H}\to\mathrm{Z}\mathrm{Z}^*\to 4j$) or have too low $\mathcal{B}$'s  (\eg\ $\mathrm{H}\to\mathrm{Z}\mathrm{Z}^*\to4\ell$) and thereby have zero expected counts for any realistic integrated luminosity. The generator-level background cross sections in Table~\ref{tab:Sigma_Higgs} are indicatively quoted
without ISR to avoid artificial enhancements of their values due to radiative-returns to the Z pole, which can be easily removed experimentally (\eg\ tagging the ISR photon and/or imposing requirements on the total energy of the event).
The last column lists the indicative signal-over-background ($S/B$) expected for the dominant (irreducible) background of each channel, at the generator level without any analysis cuts. Three broad categories can be identified:
\begin{description}
\item i) Final states with pairs of jets or tau leptons, with very large backgrounds leading to $S/B\approx 10^{-7}$--$10^{-5}$, except for the $\mathrm{H}\to gg$ case for which no actual physical background exists ($Z^*,\gamma^*$ do not couple to gluons), but for an experimental misidentification probability of light-quarks for gluons that we take as 1\% (Table~\ref{tab:reco_perf});
\item ii) Final states from intermediate $\mathrm{W}\mathrm{W}^*$ decays, with $S/B\approx 10^{-3}$;
\item iii) Final states from intermediate $\mathrm{Z}\mathrm{Z}^*$ decays with $S/B\approx 10^{-2}$, but very small signal cross sections.
\end{description}
In addition, the last row of the table lists the Higgs diphoton decay mode (discovery channel at the LHC) that suffers from both, a tiny signal cross section and 8 orders-of-magnitude larger backgrounds. A swift analysis of this table allows one to identify two channels with some potentiality in terms of statistical significances, $\mathrm{H}\to gg$ and $\mathrm{H}\to\mathrm{W}\mathrm{W}^*\to \ell\nu\;2j$, which both feature $\sim$25-ab cross sections and $S/B\approx 10^{-3}$.


\begin{table}[htbp!]
\centering
\caption{Cross sections (including ISR and $\delta_{\sqrts} = 4.1$\,MeV) times branching fractions ($\mathcal{B}$) for 11 final states in $\epem\to\mathrm{H}(XX)$ signal processes and associated dominant $\epem\to XX$ backgrounds (without ISR), and ratio of signal-over-background for each channel before any analysis cuts  (the digluon S/B quoted assumes a light-$q\to g$ mistagging rate of 1\%).}
\begin{tabular}{lcc|lc|c}\hline
Higgs decay channel & $\mathcal{B}$ & $\sigma\times\mathcal{B}$ & Irreducible background & $\sigma$ & $S/B$\\\hline
$\epem \to \mathrm{H} \to \bbbar$      & 58.2\% &  164 ab              & $\epem \to \bbbar$ & 19 pb & $\mathcal{O}(10^{-5})$ \\
$\epem \to \mathrm{H} \to gg$          &  8.2\% &  23 ab               & $\epem \to \qqbar$ & 61 pb & $\mathcal{O}(10^{-3})$ \\
$\epem \to \mathrm{H} \to \tau\tau$    &  6.3\% &  18 ab & $\epem \to \tau\tau$ & 10 pb & $\mathcal{O}(10^{-6})$ \\
$\epem \to \mathrm{H} \to \ccbar$      &  2.9\% &   8.2 ab & $\epem \to \ccbar$ & 22 pb & $\mathcal{O}(10^{-7})$ \\\hline
$\epem \to \mathrm{H} \to \mathrm{W}\mathrm{W}^* \to \ell\nu\;2j$ & 21.4\%$\times$67.6\%$\times$32.4\%$\times$2  & 26.5 ab & $\epem \to \mathrm{W}\mathrm{W}^* \to \ell\nu\;2j$ & 23 fb  & $\mathcal{O}(10^{-3})$ \\
$\epem \to \mathrm{H} \to \mathrm{W}\mathrm{W}^* \to 2\ell\;2\nu$ & 21.4\%$\times$32.4\%$\times$32.4\% & 6.4 ab  & $\epem \to \mathrm{W}\mathrm{W}^* \to 2\ell\;2\nu$ &  5.6 fb & $\mathcal{O}(10^{-3})$\\
$\epem \to \mathrm{H} \to \mathrm{W}\mathrm{W}^* \to 4j$          & 21.4\%$\times$67.6\%$\times$67.6\% & 27.6 ab & $\epem \to \mathrm{W}\mathrm{W}^* \to 4j$ & 24 fb & $\mathcal{O}(10^{-3})$ \\\hline
$\epem \to \mathrm{H} \to \mathrm{Z}\mathrm{Z}^* \to 2j\;2\nu$    &  2.6\%$\times$70\%$\times$20\%$\times$2 & 2 ab  & $\epem \to \mathrm{Z}\mathrm{Z}^* \to 2j\;2\nu$ &  273 ab & $\mathcal{O}(10^{-2})$ \\
$\epem \to \mathrm{H} \to \mathrm{Z}\mathrm{Z}^* \to 2\ell\;2j$   &  2.6\%$\times$70\%$\times$10\%$\times$2 & 1 ab & $\epem \to \mathrm{Z}\mathrm{Z}^* \to 2\ell\;2j$ &  136 ab & $\mathcal{O}(10^{-2})$ \\
$\epem \to \mathrm{H} \to \mathrm{Z}\mathrm{Z}^* \to 2\ell\;2\nu$ &  2.6\%$\times$20\%$\times$10\%$\times$2 & 0.3 ab  & $\epem \to \mathrm{Z}\mathrm{Z}^* \to 2\ell\;2\nu$ &  39 ab & $\mathcal{O}(10^{-2})$ \\\hline
$\epem \to \mathrm{H} \to \gaga$          &   0.23\% & 0.65 ab & $\epem \to \gaga$          & 79 pb & $\mathcal{O}(10^{-8})$ \\
 \hline
\end{tabular}
\label{tab:Sigma_Higgs}
\end{table}

 It is worth noting that the background cross sections computed with \pythia~8 for two-particle final states ($\epem\to\qqbar,\ccbar, \bbbar,\tau\tau,\gaga$) are found consistent with those obtained running alternative calculators, such as \MG~\cite{Alwall:2014hca,Hirschi:2015iia}, but that those for 4-fermion processes with intermediate $\mathrm{W}\mathrm{W}^*$ and $\mathrm{Z}\mathrm{Z}^*$ are prone to ambiguities in the internal definition of the contributing diagrams, and the ISR treatment, and are not always numerically compatible among them. We trust that such differences will not significantly alter our final results, given that the applied multivariate analysis will remove most non-signal-like topologies, but a dedicated study of 4-fermion backgrounds with an alternative MC generator (such as \whizard~\cite{Kilian:2007gr} or \kkmc~\cite{Arbuzov:2020coe}) is left for a forthcoming work. In this context, a few of the quoted background diboson cross sections in Table~\ref{tab:Sigma_Higgs} should be just taken as indicative of the order-of-magnitude irreducible contributions expected for the corresponding Higgs decay.

\vspace{-0.2cm}
\section{Event reconstruction and preselection}
\label{sec:presel}

Signal and background events are generated, showered, and decayed with \pythia~8 (v2.26). Initial state radiation is activated for all backgrounds, and the signal cross section samples are scaled to the ISR-plus-energy-spread benchmark point discussed in Section~\ref{sec:intro}. 
A detector polar angle acceptance of $5^\circ \gtrsim \theta \gtrsim 175^\circ$ is assumed for all final-state particles (defined as those with lifetime $c\tau_0>10$\,mm). The \fastjet\ package~\cite{Cacciari:2011ma} is used to reconstruct all jets using the $k_\mathrm{T}$ algorithm~\cite{Catani:1993hr,Ellis:1993tq} in its exclusive variant that clusterizes all hadrons in the event into a prefixed number $N_j=2,4$ of jets (the $N_j$ choice depends on the particular final state aimed at, \eg\ $\mathrm{H}\to\qqbar\to 2j$, $\mathrm{H}\to\mathrm{W}\mathrm{W}^*,\mathrm{Z}\mathrm{Z}^*\to\,2j+\ell/\nu$, or $\mathrm{H}\to\mathrm{W}\mathrm{W}^*\to\,4j$). Whenever photons or charged leptons are required to be isolated, standard criteria are applied: the sum of all particles energies must be below 1\,GeV within a radius $\Delta R=0.25$ around the $\gamma$ or $\ell^\pm$ direction. Neutrinos and particles beyond the angular acceptance are added to the missing energy ($E_\mathrm{miss}$) 4-vector. The impact of detector (in)efficiencies on the reconstruction of relevant final states is implemented in a simplified manner, according to the performances listed in Table~\ref{tab:reco_perf}. 
\begin{table}[htbp!]
\centering
\caption{Bottom ($b$), charm ($c$), and light ($uds$) quarks, gluon ($g$), tau lepton ($\tau_\text{had}$, hadronically decaying), and photon/electron (mis)reconstruction performances assumed in this study.}
\resizebox{\textwidth}{!}
{
\begin{tabular}{llllll}\hline
   & $b$ jets & $c$ jets & gluon jets & $\tauhad$ & $\gamma,e^\pm$\\\hline
reco/tagging efficiency ($\varepsilon_i$)  
& 80\%    & 70\%    & 70\%  & 80\%  & 100\% \\
mistagging rates ($\varepsilon^\mathrm{mistag}_{j\to i}$) 
& 1\% (for $c$ jet) & 5\% (for $b$ jet) &  1\% (for $uds$ jets) & $\sim$0\% (for $b,c$-jets) & 0.01\% ($e^\pm$ for $\gamma$)\\
& 0.01\% (for $udsg$ jets) & 0.1\% (for $udsg$ jets) & 0.001,0.01\% (for $b,c$-jets) & $\sim$0\% (for $udsg$ jets) & \\\hline
\end{tabular}
}
\label{tab:reco_perf}
\end{table}
The (mis)tagging jet-flavour performances are beyond the current state-of-the-art reached at the LHC today, but reasonably achievable in the ``clean'' environment of $\epem$ collisions with dedicated high-precision FCC-ee detectors after various years of operation at the Z pole and HZ energies. More details on the various jet working points assumed are provided in the next section.
We note that since the analysis boils down to basically just counting the number of events sharing a given predefined final state, any detector resolution/smearing effects on kinematic properties of the reconstructed objects (jets, $\ell^\pm$, $\gamma$,...) impact identically signals and backgrounds, will be very well controlled comparing real data and simulations, and can be accounted for here just through a (small) assigned systematic uncertainty on the final yields when computing the final statistical significance of each channel.

In Table~\ref{tab:final_states}, we list the criteria applied to all signal and backgrounds events aiming at a first preselection of final-state topologies consistent with each considered Higgs decay channel. The goal of this first set of cuts is to remove reducible backgrounds as much as possible, while keeping the largest possible signal cross section.  For the $\mathrm{H}\to\tau\tau$ channel, we consider only the fully hadronic ($\tauhad\tauhad$) decay, which is $0.65\!\cdot\!0.65/(0.35\!\cdot\!0.35)\approx 3.5$ times more probable than the fully leptonic one $\mathrm{H}\to\taulep\taulep$ (that has thereby a negligible number of signal counts expected after cuts). The last column quotes the approximate percentage of cross section signal retained by the chosen criteria.
\begin{table}[htpb!]
\caption{Minimal event final-state definition for each considered Higgs decay channel, and associated preselection efficiency (after acceptance, and reconstruction (in)efficiencies of Table~\ref{tab:reco_perf}). The $\ell^\pm$ symbol indicates $\mathrm{e}^\pm, \mu^\pm, \taulep^\pm$ charged leptons.
\label{tab:final_states}}
\centering
{
\begin{tabular}{llc}
\hline
Target Higgs decay & Final state definition & Signal presel.\ efficiency \\\hline
$\mathrm{H}\to\bbbar$ & 2 (excl.) jets, 1 $b$-tagged jet, no $\tauhad$ & 80\% \\
$\mathrm{H}\to gg$    & 2 (excl.) gluon-tagged jets, 0 isolated $\ell^\pm$ & 50\% \\
$\mathrm{H}\to\tauhad\tauhad$ & Exactly 2 $\tauhad$, 0 isolated $\ell^\pm$ & 65\%\\
$\mathrm{H}\to\ccbar$ & 2 (excl.) jets, 1 $c$-tagged jet, no $\tauhad$ & 70\% \\\hline
$\mathrm{H}\to\mathrm{W}\mathrm{W}^*\to\ell\nu 2j$ & 1 isolated $\ell^\pm$, $E_\mathrm{miss}>2$ GeV, 2 (excl.) jets & $\sim$100\%\\
$\mathrm{H}\to\mathrm{W}\mathrm{W}^*\to 2\ell 2\nu$ & 2 isolated  opp.-charge $\ell^\pm$, $E_\mathrm{miss}>2$ GeV, 0 non-isol.\,$\ell^\pm$, 0 charged hadrons   & $\sim$100\%\\
$\mathrm{H}\to\mathrm{W}\mathrm{W}^*\to 4j$ & 4 (excl.) jets, $\geq 1$ $c$-tag jets, 0 $b$-,$g$-tag jets; & 70\%\\
& jets with $m_{j1j2}\approx m_\mathrm{W}$ not both $c$-tagged, 0 $\tauhad$, 0 isolated $\ell^\pm$ & \\\hline
$\mathrm{H}\to\mathrm{Z}\mathrm{Z}^*\to 2j 2\nu$ & 2 (excl.) jets, $E_\mathrm{miss} > 30$ GeV, 0 isolated $\ell^\pm$, 0 $\tauhad$ & $\sim$100\%\\
$\mathrm{H}\to\mathrm{Z}\mathrm{Z}^*\to 2\ell 2j$ & 2 isolated opposite-charge $\ell^\pm$, 2 (excl.) jets, 0 $\tauhad$ & $\sim$100\%\\
$\mathrm{H}\to\mathrm{Z}\mathrm{Z}^*\to 2\ell 2\nu$ & 2 isolated opp.-charge $\ell^\pm$, $E_\mathrm{miss}>2$ GeV, 0 non-isol.\,$\ell^\pm$, 0 charged hadrons & $\sim$100\%\\\hline
$\mathrm{H}\to\gaga$ & 2 (excl.) isolated photons & $\sim$100\%\\\hline
\end{tabular}
}
\end{table}
 
\section{Multivariate analysis (MVA) per channel}
\label{sec:MVA}

For each reconstructed event of all generated MC samples passing the aforementioned preselection criteria per target Higgs channel, we define $\mathcal{O}(50)$ variables for single and combined ($n$-wise) physics objects (jets, charged leptons, photons, neutrinos), as well as for global event properties, in order to provide as much information as possible to a subsequent MVA used to discriminate signal and the remaining backgrounds.
The defined variables include kinematic components $(\pt,\eta,\phi,E)$, charge, mass (invariant and transverse),... for each single object ---as well the same quantities for sums and differences of 4-momenta of selected $n$-wise objects combinations---, the maximum and minimum values of $\pt^{i(ij)},\,\eta^{i,(ij)},\phi^{i,(ij)}$, $m_{ij}$,... in the event for all (pairs of) objects $i$ ($ij$), as well as quantities associated with global event topologies (sphericity, linearity, aplanarity, thrust max/min,...). Angular information is particularly useful in diboson channels with decay leptons in order to separate final states coming through the spin-0 Higgs resonance or proceeding through $t$-channel processes or via the spin-1 $s$-channel continuum and/or $\mathrm{Z}^*,\gamma^*,\mathrm{W}^\pm$ decays. For such cases, angular discrimination variables based on the Matrix Element Likelihood Analysis (MELA)~\cite{Chatrchyan:2012jja} are also defined and incorporated into the MVA.
We used the TMVA framework~\cite{Hocker:2007ht}  to train and test boosted decision-tree (BDT) classifiers
in order to provide statistical discrimination between each Higgs decay channel and all 
relevant background final states,
and maximize the signal significance. Examples of the BDT variables used for a particular channel ($\mathrm{H}\to\mathrm{W}\mathrm{W}^*\to\ell\nu 2j$) are shown in Fig.~\ref{fig:BDT} (right) later, as well as listed with their individual relative weights in the final signal significance in Table~\ref{tab:BDT_vars}.




Table~\ref{tab:N_evts_per_channel} lists the number of signal and background(s) events expected after preselection and BDT output cuts, for 9 different final states. 
We omit the $\mathrm{H}\to\gaga,\ccbar$ channels from the table given that they are fully swamped by backgrounds and have a negligible statistical significance. The first observation is that except for the $\mathrm{H}\to\bbbar$ decay, which is anyway overwhelmed by the continuum background, the final number of signal events is (well) below 100 counts for each individual channel and that the remaining backgrounds counts are orders-of-magnitude larger. Therefore, the leading uncertainty of the signal will be of statistical nature, and evidence of any excess will rely on an accurate control of the background systematic uncertainties (which must be well below the statistical ones). Among the listed channels, we observe that $\mathrm{H}\to gg$ and $\mathrm{H}(\mathrm{W}\mathrm{W}^*) \to \ell\nu\;2j$ feature the largest $S/\sqrt{B}$ significances\footnote{The actual significance per channel is computed for a single-bin counting experiment using a profile likelihood approach, as explained in Section~\ref{sec:results}, but it is numerically consistent with this standard naive estimation.}, and are discussed in more detail in dedicated subsections below. The $\mathrm{H}\to\bbbar$ channel suffers from a very large irreducible background, the MVA is unable to improve the rejection of the $\epem\to\bbbar$ continuum much beyond the preselection result, and the final statistical significance remains very low ($S/\sqrt{B}\approx 0.12$). Although orders-of-magnitude smaller, we also quote the number of misidentified $\epem\to\ccbar,\qqbar$ background events expected for this channel, so as to assess the potential contamination from such processes 
if the mistagging points assumed in Table~\ref{tab:reco_perf} are changed. The $\mathrm{H}\to\tauhad\tauhad$ decay mode (as well as, similarly, the $\mathrm{H}\to\ccbar$ one not listed) suffers from very low signal counts and a daunting continuum background that yields a negligible statistical significance ($S/\sqrt{B}\approx 0.02$). Among $\mathrm{H}\to\mathrm{W}\mathrm{W}^*$ final states, the fully leptonic one ($2\ell2\nu$) features the smallest branching fraction, and thereby very low final signal counts. For the two others, lepton+jets ($\ell\nu\;2j$) and fully hadronic ($4j$) decays, although they have the same branching fraction, only the former can take advantage of background removal by exploiting the different $\mathrm{W}^\pm\to\ell^\pm$ decay lepton polarizations for signal and background processes, as explained below. Finally, Table~\ref{tab:N_evts_per_channel} shows that the $\mathrm{H}\to\mathrm{Z}\mathrm{Z}^*$ final states will have less than $\sim$10 signal events expected after cuts, over much larger backgrounds, and appear statistically marginal in terms of signal significance.

\begin{table}[htpb!]
\caption{Number of reconstructed events expected after preselection N(presel.) and BDT output N(MVA) cuts, for $s$-channel Higgs decay modes and associated dominant backgrounds in $\epem$ collisions at $\sqrts = m_\mathrm{H}$ ($\delta_{\sqrts} = 4.1$\,MeV and $\LumiInt = 10$\,ab$^{-1}$).
\label{tab:N_evts_per_channel}}
\centering
\resizebox{\textwidth}{!}
{
\begin{tabular}{lcc|lcc|lcc}\hline
Channel & N(presel.) & N(MVA) & Channel & N(presel.) & N(MVA) & Channel & N(presel.) & N(MVA) \\\hline
\rowcolor{Gray}
$\mathrm{H}\to\bbbar$ & 1320  & 1220 & 
$\mathrm{H}\to gg$ & 110 & 55 &
$\mathrm{H}\to\tauhad\tauhad$ & 48 & 13 \\
$\epem\to\bbbar$ & $1.5\cdot10^{8}$ & $1.1\cdot10^{8}$ & 
$\epem\to\qqbar$ & 61\,000 & 2400 & 
$\epem\to\tauhad\tauhad$ & $2.7\cdot10^{7}$ & $3.8\cdot10^{5}$ \\
$\epem\to\ccbar$ & $1.4\cdot10^{6}$ & $9.4\cdot10^{5}$ & 
$\epem\to\ccbar$ & 220 & $\sim$10 & 
 &  & \\ 
$\epem\to\qqbar$ & $3.0\cdot10^{4}$ & $4800$ & 
$\epem\to\bbbar$ & 20 & $\sim$1 &
 &  & \\ 
\hline
\rowcolor{Gray}
$\mathrm{H}\to\mathrm{W}\mathrm{W}^*\to\ell\nu 2j$  & 265 & 55 & 
$\mathrm{H}\to\mathrm{W}\mathrm{W}^*\to 2\ell 2\nu$ &  64 & 25 & 
$\mathrm{H}\to\mathrm{W}\mathrm{W}^*\to 4j$         & 180 & 27 \\
$\epem\to \mathrm{W}\mathrm{W}^*\to\ell\nu 2j$   & $2.3\cdot 10^5$ & 11\,000  & 
$\epem\to \mathrm{W}\mathrm{W}^*\to 2\ell 2\nu$  & $5.6\cdot 10^4$ &  7600 & 
$\epem\to \mathrm{W}\mathrm{W}^*\to 4j$          & $1.3\cdot 10^5$ &  14\,000  \\
$\epem\to \bbbar$                 & 1100 & -- & 
$\epem\to \mathrm{Z}\mathrm{Z}^*\to 2\ell 2\nu$ & 1360 & $\sim$5 &
$\epem\to \mathrm{Z}\mathrm{Z}^*\to 4j$         & $4.7\cdot 10^3$ & 20 \\
$\epem\to \ccbar,\qqbar$ & 150 & -- & 
$\epem\to \tau\tau$      & $1.2\cdot 10^7$ & -- & 
$\epem\to \bbbar,\ccbar$ & $5\cdot 10^5$ & 7\,000 \\
\hline
\rowcolor{Gray}
$\mathrm{H}\to\mathrm{Z}\mathrm{Z}^*\to 2j 2\nu$    & 21 & 11 & 
$\mathrm{H}\to\mathrm{Z}\mathrm{Z}^*\to 2\ell 2j$   & 10 &  4 &
$\mathrm{H}\to\mathrm{Z}\mathrm{Z}^*\to 2\ell 2\nu$ &  3 & 0.8 \\
$\epem\to \mathrm{Z}\mathrm{Z}^*\to 2j 2\nu$  & 2700 & 1000 & 
$\epem\to \mathrm{Z}\mathrm{Z}^*\to 2\ell 2j$ & 1000 &  500 & 
$\epem\to \mathrm{Z}\mathrm{Z}^*\to 2\ell 2\nu$ & 270 & 70 \\
$\epem\to \mathrm{W}\mathrm{W}^* \to 2j 2\nu$  & 6100 & 400 & 
$\epem\to \mathrm{W}\mathrm{W}^* \to 2\ell 2j$ & $3.3\cdot 10^4$  & $\sim$1 &
$\epem\to \mathrm{W}\mathrm{W}^*\to 2\ell 2\nu$& $3.3\cdot 10^4$  & 260 \\
$\epem\to \bbbar,\ccbar,\qqbar$ & 7000 & -- & 
$\epem\to \bbbar,\ccbar,\qqbar$ & 400 & -- & 
$\epem\to \bbbar,\ccbar,\qqbar$ & 390 & --\\
$\epem\to \tau\tau$ & 1700 & $\sim$2 & 
&  &  & 
$\epem\to \tau\tau$ &  $3\cdot 10^4$ & -- \\
\hline
\end{tabular}
}
\end{table}



\subsection*{Analysis of \texorpdfstring{$\epem \to \mathrm{H}(gg) \to jj$}{e+e- to H(gg)}}
\label{sec:Hgg}

\begin{figure}[htpb!]
\centering
\includegraphics[width=0.495\columnwidth]{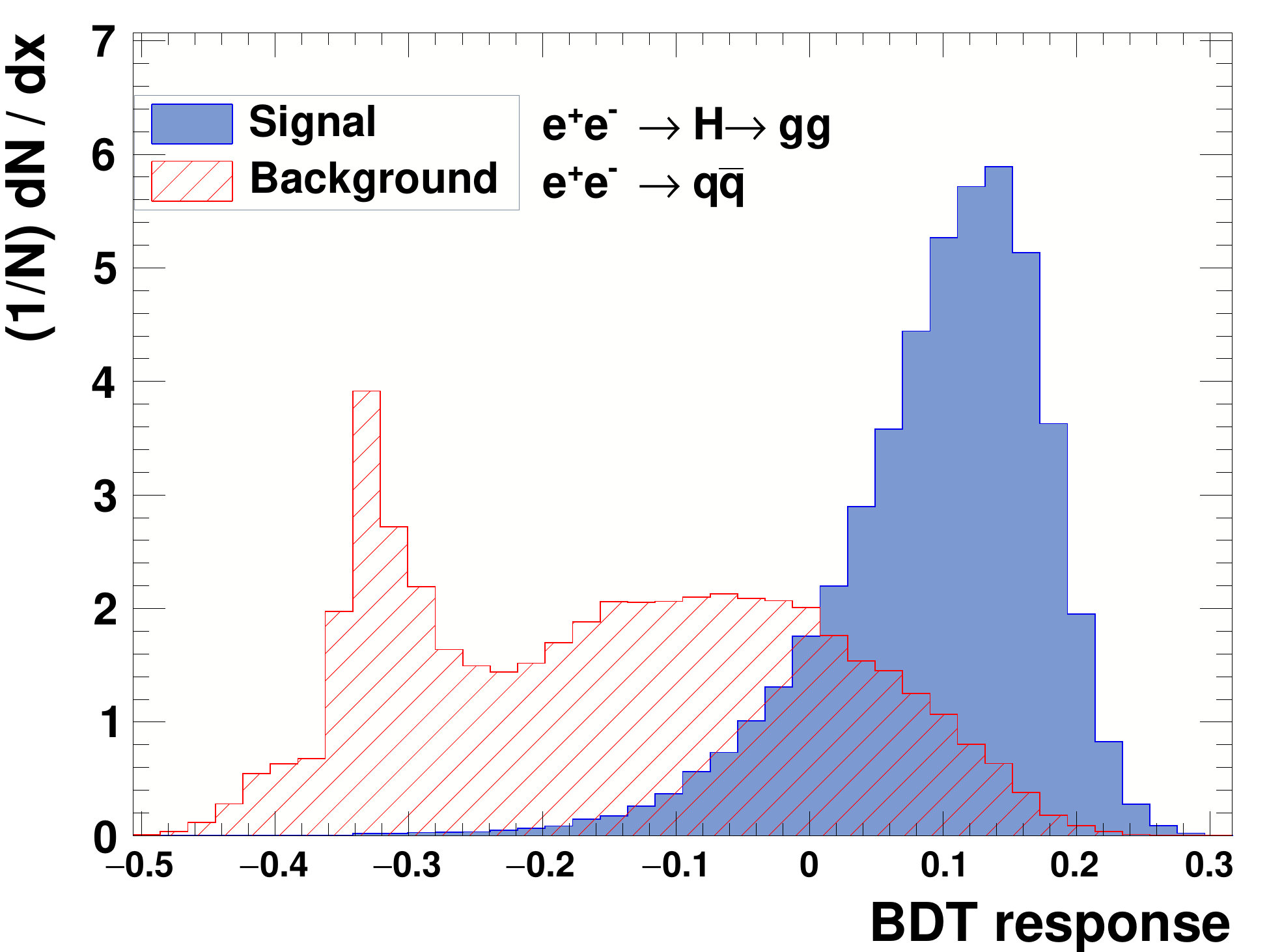}\quad
\includegraphics[width=0.485\columnwidth]{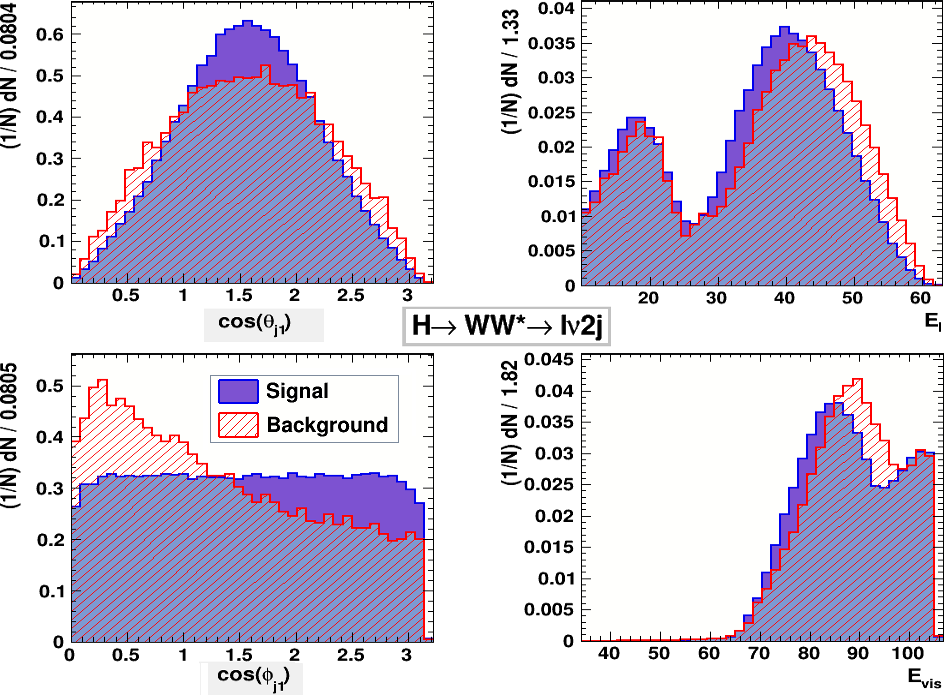}
\caption{Left: Example of normalized BDT response distributions for signal and backgrounds in the $\mathrm{H}\to gg$ channel. Right: Examples of a few of the most discriminating (normalized) BDT variables of the $\mathrm{H}\to \mathrm{W}\mathrm{W}^* \to \ell\nu\;2j$ analysis. 
\label{fig:BDT}}
\end{figure}

At face value, the digluon decay is a very promising signal channel as it has the third most abundant Higgs branching fraction ($\mathcal{B}(\mathrm{H}\to gg) = 8.2\%$), and has no irreducible physical background because Z and $\gamma$ bosons do not couple to gluons. However, the production of light quark ($uds$) pairs in the much more abundant $\epem\to\mathrm{Z}^*,\gamma^*\to\qqbar$ process (with cross sections million times larger than that of the signal, Table~\ref{tab:Sigma_Higgs}), jeopardizes the observation of $\mathrm{H}\to gg$ because experimentally separating jets issuing from the showering and hadronization of light-quarks and from gluons is not perfect\footnote{Separating heavy-quark ($c$, $b$) from gluon jets is easier given the presence of \textit{single} bottom/charm hadrons in the former, which  in the gluon case only appear in pairs, through gluon splitting, with suppressed probabilities $g_{g\to\ccbar,\bbbar}\approx 3\%,0.3\%$ at the Z mass~\cite{Zyla:2020zbs}. Although somewhat arbitrary, we have considered tiny but finite mistagging rates $\varepsilon^\mathrm{mistag}_{b,c\to g} =0.001,0.01\%$ to be able to quantify the impact from such sources (rescaling them, if needed) in the number of background events identified as digluons (Table~\ref{tab:N_evts_per_channel}).}. An illustrative case would be the emission of a very hard gluon from each one of the $\mathrm{Z}^*\to\qqbar$ quarks, that could mimic the Higgs digluon final state. Fortunately, in the last years there has been tremendous progress on quark-gluon tagging studies exploiting jet substructure properties with machine learning techniques~\cite{Larkoski:2017jix}. The latest LHC results reach $\varepsilon_g\approx 60\%$ gluon efficiencies with $\varepsilon^\mathrm{mistag}_{q\to g}\approx 10\%$ false positive rates using advanced multivariate analyses~\cite{Kasieczka:2018lwf,Filipek:2021qbe}, or $\varepsilon^\mathrm{mistag}_{q\to g}\approx 7\%$~\cite{Khosa:2021cyk} further exploiting Lund jet plane information~\cite{Dreyer:2018nbf}. Reaching mistagging rates down to $\varepsilon^\mathrm{mistag}_{q\to g}\approx 1\%$, while keeping large gluon reconstruction efficiencies, appears feasible in the clean and kinematically constrained QCD environment of future $\epem$ machines, in particular taking advantage of the very large samples of $\mathrm{Z}\to\qqbar(g)$ events at the Z pole, and the $\mathcal{O}(10^5)$ $\mathrm{H}\to gg$ events collected during the $\epem\to\mathrm{Z}\mathrm{H}$ runs, available for dedicated studies of the different colour, radiation, spin, charge, hadronization properties of quark and gluon jets~\cite{Skands:2016bxb,Anderle:2017qwx,dEnterria:2019jfn}. The addition of advanced hadron identification capabilities to the FCC-ee detectors for dedicated flavour~\cite{Wilkinson:2021ehf} (and QCD) studies, will further reduce the parton-to-hadron fragmentation uncertainties~\cite{DdE_jets}. Our assigned (mis)reconstruction jet working point for this channel is $(\varepsilon_g,\varepsilon^\mathrm{mistag}_{q\to g})=(70\%,1\%)$, which leads to a $10^{-4}$ background rejection factor when requiring two gluon-tagged exclusive jets in the event. The corresponding number of events expected in 10\,ab$^{-1}$ for signal and background, after acceptance and efficiency preselections, are 110 and $\sim$61\,000 respectively (Table~\ref{tab:N_evts_per_channel}). The subsequent MVA is performed removing beforehand any jet variable that may have been potentially used to define the light-$q$/gluon separation, and which is therefore de facto already accounted for by the chosen preselection (mis)tagging efficiency. An analysis of the BDT response (Fig.~\ref{fig:BDT}, left) indicates a maximum significance reached for a BDT output cut that further reduces the background by a factor $\times 0.06$ while only losing 50\% of the signal. The final statistical significance reached for this channel is approximately given by $S/\sqrt{B} = 55/\sqrt{2400} = 1.1\sigma$ per FCC-ee IP per year. 
\vspace{-0.4cm}
\subsection*{Analysis of \texorpdfstring{$\epem \to \mathrm{H}(\mathrm{W}\mathrm{W}^*) \to \ell\nu\;2j$}{e+e- to H(WW) to lnu2j} }
\label{sec:HWWlj}

The event signature of the $\mathrm{H}(\mathrm{W}\mathrm{W}^*) \to \ell\nu\;2j$ signal is one isolated charged lepton, missing energy from the neutrino, and two exclusive jets. In principle, such a final state can be present in multiple reducible backgrounds (Table~\ref{tab:N_evts_per_channel}), but the MVA study allows to remove basically all of them, leaving just a fraction of the $\epem \to \mathrm{W}\mathrm{W}^* \to \ell\nu\;2j$ continuum. Table~\ref{tab:BDT_vars} lists the BDT variables used in the analysis, together with their relative weight in the final signal significance for this channel. Apart from blindly running the MVA, it is instructive to show the impact of different kinematic cuts to get rid of reducible backgrounds. Thus, for example, a significant fraction of $\epem\to\qqbar,\ccbar,\bbbar$ events can be eliminated by requiring \eg: $E_{j1,j2} < 52,45$ GeV, $m_{\mathrm{W}(\ell\nu)} > 12$ GeV, $E_{\ell} > 10$ GeV, $E_\mathrm{miss} >$ 20 GeV. The additional requirement on the mass of the missing 4-momentum vector $m_\mathrm{miss} < 3$ GeV further discards many $\epem\to\tau\tau$ events. 
\begin{table}[htpb!]
\centering
\caption{Indicative list of BDT variables used in the $\mathrm{H} \to \mathrm{W}\mathrm{W}^* \to \ell\nu\;2j$ analysis, with their relative weight in the statistical significance for this channel.
\label{tab:BDT_vars}}
\begin{tabular}{llllllllll}\hline
$\cos\theta_{j1}$ & $E_{\ell}$  & $\pt(jj)$ & $\cos\phi_{j1}$ & $m_\mathrm{miss}$ & $E_\mathrm{visible}$ & $\pt^{\ell}$ & $E_\mathrm{miss}$ & $\pt(jj{\ell})$ & $\cos\theta^*$ \\
0.0446 & 0.0417 & 0.0409 & 0.0398 & 0.0341 & 0.0328 & 0.0308 & 0.03015 & 0.02726 & 0.02626 \\\hline
$\eta_\mathrm{miss}$ & $\eta_{j1}$ & $\cos\theta_{j2}$ & $\Delta\phi_{jj}$ & $m_{_{\mathrm{T,miss}}}$ & $m_\mathrm{W\,offsh.}$ & $E_{j,\mathrm{min}}$ & $\Delta R_\mathrm{min,j\ell}$ & $\min\Delta\eta_{j\ell}$  & $\pt^{j1}$ \\
0.0255  & 0.0238 & 0.0220 & 0.0215 & 0.0212 & 0.0212 & 0.0205 & 0.0204 & 0.0192 & 0.0189 \\\hline
$\max\cos(\ell j)$ & $\eta_{\ell}$ & $m(\ell\nu)$ & $\min\cos(\ell j)$ & $\max\Delta\eta_{jj}$  & $m_\mathrm{W\,shell}$ & $\mT(\ell j_1)$ & $\mT(jj\ell)$ & $m(\ell j_1)$ & $m_{j2}$ \\
0.0189 & 0.0182 & 0.0179 & 0.0176 & 0.0165 & 0.0160 & 0.0160 & 0.0160 & 0.0156 & 0.0147 \\\hline
$\cos\phi_{j1,j2}$ & $\pt^{j2}$ & $\Delta R_\mathrm{max,j\ell}$ & $\eta_{j2}$ & lin.spher. & $m_{j1}$ & $\pt(\ell j2)$ & $\Delta\theta_{jj}$ & $\mT(jj)$ & $\Delta R_{jj}$ \\
0.0140 & 0.0136 & 0.0136 & 0.0136 & 0.0136 & 0.0134 & 0.0134 & 0.0132 & 0.0131  & 0.0127 \\\hline
 $E_{j,\mathrm{max}}$ & $\mT({\ell}j_2)$ & sphericity & $\pt(\ell j1)$ & $\min\Delta\phi_{j\ell}$ & $\mathrm{E_{isol}}$ & aplanarity & $\max\Delta\phi_{j\ell}$  & $\phi_{j_1}$ & $m(jj\ell)$ \\
0.0125 & 0.0121 & 0.0116 & 0.0103 & 0.0102 & 0.00998 & 0.00927 & 0.00914 & 0.00894  &0.00764 \\\hline
$m({\ell}j_2)$ & $m_{jj}$ & $\phi_{j_2}$ & lin.aplan. & $\phi_{\ell}$ & $\cos\phi^*$ & \multicolumn{4}{c}{others ($R_\mathrm{min}$, $\eta_{\ell}$, \dots)} \\
  0.00680 & 0.00641 & 0.00565 & 0.00514 & 0.00512 & 0.00471 &\multicolumn{4}{c}{$<0.001$} \\\hline
\end{tabular}
\end{table}

The remaining background is dominated by the $\mathrm{W}\mathrm{W}^*$ continuum that can then be reduced by exploiting, among others, the different $\mathrm{W}^\pm$ polarizations for signal and background processes. The signal decay $\mathrm{H} \to \mathrm{W}^+\mathrm{W}^-$ is that of a scalar to a pair of distinguishable spin-1 bosons. The subsequent W bosons decays maximally violate chirality: a $\mathrm{W}^+$ ($\mathrm{W}^-$) boson preferentially emits a $\ell^+$ ($\ell^-$) along (against) its spin direction. The anticorrelation between the $\mathrm{W}^\pm$ polarizations expected in spin-zero Higgs decays is transferred into a correlation between the momenta of the charged leptons in their decays that manifests itself in the distributions of relative $\ell^\pm$ polar angles, and a preference for a small azimuthal angle ($\phi$) between the $\ell^+\ell^-$ pair. Such angular correlations of the emitted charged leptons are encoded into the MELA variables exploited by the ATLAS and CMS collaborations to separate Higgs decays from $\mathrm{W}^+\mathrm{W}^-$ backgrounds in their original searches~\cite{Aad:2012tfa,Chatrchyan:2012ufa}. Examples of discriminating BDT variables distributions for signal and backgrounds are shown in Fig.~\ref{fig:BDT} (right).
Applying an appropriate cut on the BDT response output, keeps a 58\% efficiency on signal, while removing 80\% of the continuum background. The final statistical significance of this final state is of the order of $S/\sqrt{B} = 55/\sqrt{11\,000}\approx 0.5\sigma$ 
per FCC-ee IP per year.

\section{Beam monochromatization, expected signal significance and \texorpdfstring{$y_\mathrm{e}$}{y(e)} constraints}
\label{sec:results}

Table~\ref{tab:significances} lists the statistical significances, in units of std.\ deviations $\sigma$, for each individual $s$-channel Higgs decay channel studied here, for our baseline $(\delta_{\sqrts},\LumiInt) = (4.1\,\mathrm{MeV},10\,\mathrm{ab}^{-1})$ monochromatization assumption. The combined final significance, and associated 95\% CL upper limit, are calculated considering a multibin counting experiment with a profile likelihood for hypothesis test and confidence interval, using the \textsc{RooStats} statistical package~\cite{Moneta:2010pm}. We have considered $10^{-4}$ fractional systematic uncertainties\footnote{A detailed description of the systematic studies and detector requirements needed to achieve such uncertainties for each of the Higgs final states is beyond the scope of this essay, and will be part of the outcome of the forthcoming FCC feasibility study.} for the backgrounds, consistent with the expected experimental precision aimed at FCC-ee~\cite{Blondel:2019jmp}.
The final combined significance is $1.3\sigma$, which is also very close to the naive quadratic sum of individual $S/\sqrt{B}$ values per channel. Such a result is equivalent to setting a 95\% CL upper limit of 2.6 times the SM Higgs $s$-channel cross section, per FCC-ee IP and per year. Since the cross section depends on the square of the electron Yukawa, $\sigma_{\epem\to\mathrm{H}}\propto y_\mathrm{e}^2$, this corresponds to placing an upper bound on the coupling at $\sqrt{2.6}=1.6$ times the SM value, \ie\ $|y_\mathrm{e}|<1.6|y^\mathrm{\textsc{sm}}_\mathrm{e}|$ (95\% CL).
\begin{table}[htbp!]
\centering
\caption{Individual significances (in std.\ deviations $\sigma$) expected per decay channel for $s$-channel Higgs boson production in $\epem$ collisions at FCC-ee for $\LumiInt = 10$\,ab$^{-1}$ and $\delta_{\sqrts}=4.1$\,MeV. The last column quotes the combined significance.
\label{tab:final_signifs}}
\begin{tabular}{ccccc|c}\hline
$\mathrm{H}\to gg$ & $\mathrm{H}\to\mathrm{W}\mathrm{W}^* \to \ell\nu\;2j;\,2\ell\;2\nu;\,4j$ & $\mathrm{H}\to\mathrm{Z}\mathrm{Z}^* \to 2j\;2\nu;\, 2\ell\;2j;\,2\ell\;2\nu$ & $\mathrm{H}\to\bbbar$ & $\mathrm{H}\to\tauhad\tauhad;\,\ccbar;\,\gaga$ & Combined \\\hline
 1.1$\sigma$ & $(0.53\otimes0.34\otimes0.13)\sigma$ & $(0.32\otimes0.18\otimes0.05)\sigma$ & 0.13$\sigma$ & $<0.02\sigma$  & 1.3$\sigma$ \\
\hline
\end{tabular}
\label{tab:significances}
\end{table}

\begin{figure}[htpb!]
\centering
\includegraphics[width=0.5\columnwidth]{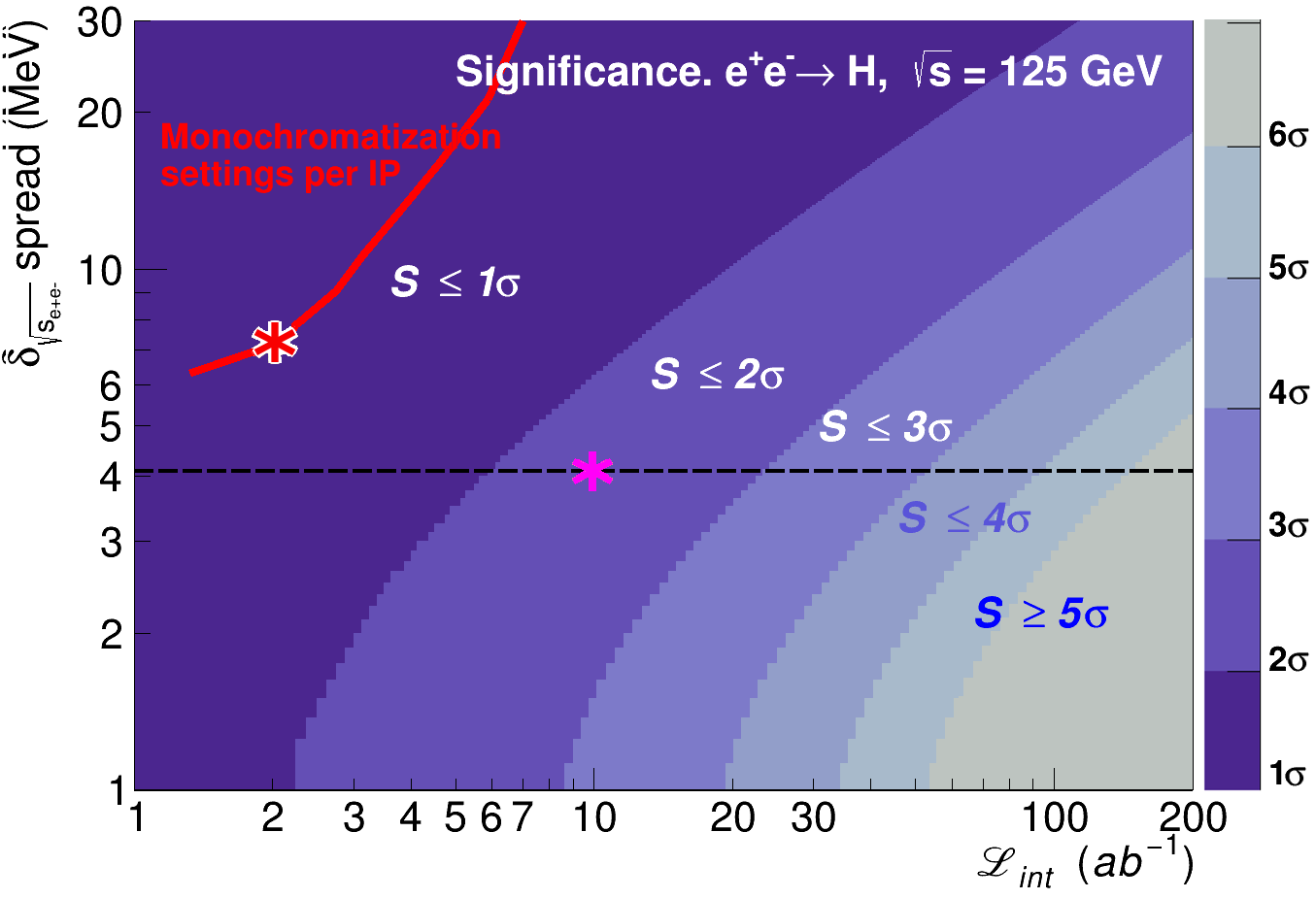}\hspace{0.1cm}
\includegraphics[width=0.48\columnwidth]{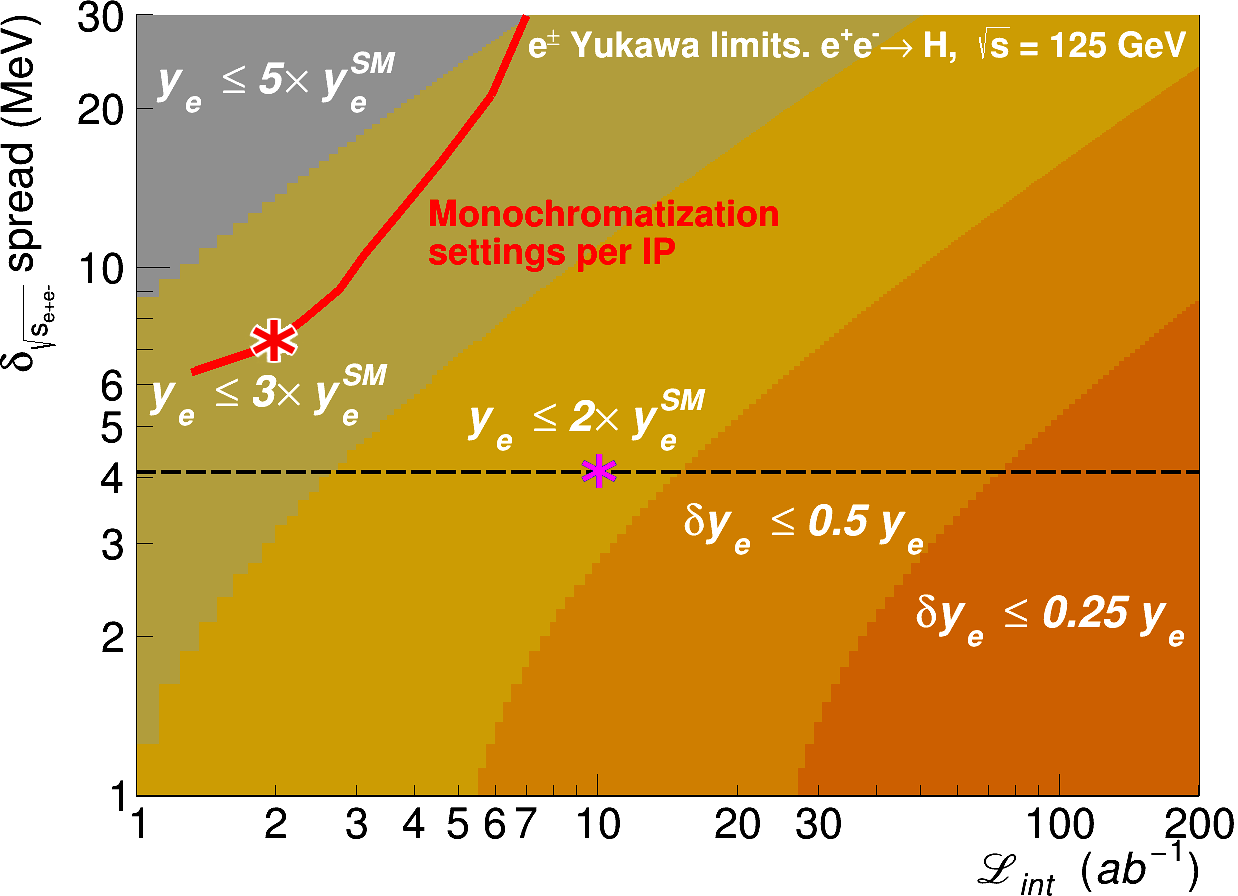}
\caption{Left: Significance contours (in std.\ dev. units  $\sigma$) in the \com\ energy spread vs.\ integrated luminosity plane for the resonant $\sigma_{\epem\to\mathrm{H}}$ cross section at $\sqrts = m_\mathrm{H}$.
Right: Associated upper limits contours (95\% CL) on the electron Yukawa $y_\mathrm{e}$. The red curves show the range of parameters presently reached in FCC-ee monochromatization studies~\protect\cite{Zimmermann:2017tjv,ValdiviaGarcia:2019ezi}. The red star indicates the best signal strength monochromatization point in the plane (the pink star over the $\delta_{\sqrts} = \Gamma_\mathrm{H} =4.1$\,MeV dashed line, indicates the ideal baseline point assumed in our default analysis). All results are given per IP and per year.
\label{fig:signif}}
\end{figure}

The expected final significance of the $\sigma_{\epem\to\mathrm{H}}$ measurement, and associated 95\% CL limits on $|y_\mathrm{e}|$, derived for a benchmark $\delta_{\sqrts} = 4.1$\,MeV collision-energy spread and $\LumiInt = 10\,\mathrm{ab}^{-1}$ integrated luminosities, can be easily derived for any other combination of $(\delta_{\sqrts},\LumiInt)$ values achievable through beam monochromatization. Figure~\ref{fig:signif} shows the bidimensional maps for the significance of $s$-channel Higgs production (left) and the corresponding 95\% CL upper limits on the electron Yukawa (right), as a function of both parameters. The signal significance, and associated upper limits, improve with the square-root of the integrated luminosity (along the $x$ axes of both plots), and diminish for larger values $\delta_{\sqrts}$ (along the $y$ axes of the maps) following the relativistic Voigtian dependence of the signal yield on the energy spread shown in Fig.~\ref{fig:diags} (right). 

The red curves in Fig.~\ref{fig:signif} show the current expectations for the range of $(\delta_{\sqrts},\LumiInt)$ values achievable at FCC-ee with the investigated monochromatization schemes~\cite{Zimmermann:2017tjv,ValdiviaGarcia:2019ezi}. Without monochromatization, the FCC-ee natural collision-energy spread at $\sqrts = 125$\,GeV is about $\delta_{\sqrts} = 46$\,MeV due to synchrotron radiation. Its reduction to the few-MeV level desired for the $s$-channel Higgs run can be accomplished by means of monochromatization, \eg\ by introducing nonzero horizontal dispersions at the IP ($D_x^*$) of \textit{opposite sign} for the two beams in collisions without a crossing angle. The beam energy spread reduction factor is given by $\lambda = \sqrt{({D_x^*}^2\sigma_{\delta}^{2})/(\varepsilon_x\beta_x^*)+1}$, where $\beta_{x(y)}^*$ denotes the horizontal (vertical) beta function at the IP and $\varepsilon_{x(y)}$ the corresponding emittance. The need to generate a significant IP dispersion implies a change of beamline geometry in the interaction region and the use of crab cavities to compensate for the existing, or remaining, crossing angle. 
A nonzero IP dispersion leads to an increase of the transverse horizontal emittance from beamstrahlung, thereby impacting the beam luminosity. 
Optimization of the IP optics parameters ($D_x^*$, $\beta_{x,y}^*$,...) yields the corresponding red curves of Fig.~\ref{fig:signif}. For the lowest collision-energy spread achieved of $\delta_{\sqrts} = 6$\,MeV, the anticipated monochromatized luminosity per IP exceeds $10^{35}\,\mathrm{cm}^{-2}\mathrm{s}^{-1}$~\cite{ValdiviaGarcia:2019ezi}. This translates into an integrated luminosity\footnote{Conversion from luminosity ($\mathcal{L}=10^{35}\,\mathrm{cm}^{-2}\mathrm{s}^{-1}$) to integrated luminosity ($\LumiInt =1.2$\,ab$^{-1}$/year/IP) assumes 185 physics days per run with a 75\% physics efficiency~\cite{Abada:2019zxq}.} of at least 1.2\,ab$^{-1}$ per IP per year. 
One can reach larger integrated luminosities at the expense of a worse beam energy spread. The point (red star) over the red curves that has the highest signal strength today corresponds to $(\delta_{\sqrts},\LumiInt) \approx (7\,\mathrm{MeV}, 2\,\mathrm{ab}^{-1})$, to be compared to our original baseline point (pink star) over the $\delta_{\sqrts} = \Gamma_\mathrm{H} =4.1$\,MeV dashed line. For such a 7-MeV \com\ energy spread, the peak of the relativistic Voigtian distribution describing the $s$-channel cross section is located at about 1\,MeV above the mass of the Higgs boson (Fig.~\ref{fig:diags}, right). Therefore, the optimal \com\ energy of the dedicated $\epem$ run needs also to be carefully chosen to maximize the resonant cross section for any given monochromatization point. 

Compared to our baseline values (pink stars on the plots), the signal significance for the currently best monochromatization settings, $(\delta_{\sqrts},\LumiInt) \approx (7\,\mathrm{MeV}, 2\,\mathrm{ab}^{-1})$, drops to $\mathcal{S}\approx 0.4\sigma$/year/IP, and the corresponding upper bound on the $\mathrm{e}^\pm$ Yukawa becomes $y_\mathrm{e}\lesssim 2.5 y^\mathrm{\textsc{sm}}_\mathrm{e}$ (95\% CL) per year and per IP. Assuming 2 years of FCC-ee operation at the Higgs pole and combining four detectors/IPs, this would translate into a $1.2\sigma$ significance and a $y_\mathrm{e}\lesssim 1.6 y^\mathrm{\textsc{sm}}_\mathrm{e}$ limit. 
Such a result, although clearly short of an evidence for $s$-channel Higgs production, is still about 100 (30) times better~\cite{Blondel:2019yqr} than that reachable at HL-LHC (FCC-hh~\cite{Benedikt:2018csr}), and would imply setting a constraint on new physics affecting the electron-Higgs coupling above $\Lambda_{\textsc{bsm}}\gtrsim 110$~TeV.

Given that any improved analysis of the Higgs decay channels is unlikely to increase much more the final signal significance, alternative paths need to be considered in order to measure more precisely the electron Yukawa coupling at FCC-ee. The possibility of introducing beam longitudinal polarizations ($P_\mathrm{L}$) would enhance the signal by $(1+P_\mathrm{L}^2)$ and suppress backgrounds by $(1-P_\mathrm{L}^2)$, \ie\ running with $P_\mathrm{L} = 68\%$ (90\%) would increase by a factor of two (four) the statistical significance of the signal. However, for realistic longitudinal polarizations reachable at FCC-ee ($P_\mathrm{L}=20$--30\%) the gain would be insufficient and higher polarizations would significantly reduce the luminosity. The only approach seemingly left to carry out an $\epem\to \mathrm{H}$ measurement with a sensitivity reaching the SM electron Yukawa level requires improving the beam monochromatization beyond the current state-of-the-art~\cite{Zimmermann:2017tjv,ValdiviaGarcia:2019ezi}. Alternative or modified monochromatization scenarios~\cite{Zholents:1988bu,Bogomyagkov:2017tpk,Telnov:2020rxp} are being explored that however, for now, do not improve the results of the red curves shown in Fig.~\ref{fig:signif}. 


\section{Summary and outlook}
\label{sec:conclusion}

The prospects for a potential FCC-ee measurement of the direct $s$-channel Higgs boson production in $\epem$ collisions at $\sqrts = m_\mathrm{H}$ have been studied as a means to determine the Higgs Yukawa coupling of the electron ($y_\mathrm{e}$). The three main challenges of such a measurement have been discussed: (i) the need to accurately know (within MeV's) beforehand the value of the Higgs boson mass where to operate the collider, (ii) the smallness of the resonant Higgs boson cross section (few hundred ab) due to ISR and beam  energy spread ($\delta_{\sqrts}$) that requires to monochromatize the beams, \ie\ reduce $\delta_{\sqrts}$ at the few MeV scale, while still delivering large (few ab$^{-1}$) integrated luminosities $\LumiInt$, and (iii) the existence of multiple backgrounds with orders-of-magnitude larger cross sections than the Higgs signal decay channels themselves. The knowledge of $m_\mathrm{H}$ with a few MeV accuracy seems feasible at FCC-ee as per dedicated studies reported in Ref.~\cite{HiggsMassFCCee}. This present work has focused on the points (ii) and (iii) above, by performing a generator-level study that has chosen as benchmark point a baseline monochromatization scheme leading to $(\delta_{\sqrts},\LumiInt)=(4.1\,\mathrm{MeV},10\,\mathrm{ab}^{-1})$, corresponding to a peak $s$-channel cross section of $\sigma_{\epem\to\mathrm{H}} = 280$\,ab. 

Large simulated event samples of signal and associated backgrounds have been generated with the \pythia~8 Monte Carlo (MC) code for 11 Higgs boson decay channels. A simplified description of the expected experimental performances has been assumed for the reconstruction and (mis)tagging of heavy-quark ($c$, $b$) and light-quark and gluons ($udsg$) jets, photons, electrons, and hadronically decaying tau leptons. Generic preselection criteria have been defined that target the 11 Higgs boson channels, suppressing reducible backgrounds while keeping the largest fraction of the signal events. A subsequent multivariate analysis of $\mathcal{O}(50)$ kinematic and global topological variables, defined for each event, has been carried out. Boosted-Decision-Trees (BDT) classifiers have been trained on signal and background events, to maximize the signal significances for each individual channel. The most significant Higgs decay channels are found to be $\mathrm{H}\to gg$ (for a gluon efficiency of 70\% and a $uds$-for-$g$ jet mistagging rate of 1\%), and $\mathrm{H}\to\mathrm{W}\mathrm{W}^*\to\ell\nu 2j$. Combining all results, a $1.3\sigma$ signal significance can be achieved, corresponding to an upper limit on the e$^\pm$ Yukawa coupling at 1.6 times the SM value: $|y_\mathrm{e}|<1.6|y^\mathrm{\textsc{sm}}_\mathrm{e}|$ at 95\% confidence level (CL), per FCC-ee interaction point (IP) and per year. Such a bound is about $\times$100 ($\times$30) times better than that reachable at HL-LHC (FCC-hh), and can be translated into a lower limit on the energy scale of any physics beyond the SM (BSM) affecting the electron Yukawa coupling, of $\Lambda_{\textsc{bsm}} \approx v^{3/2}(\sqrt{2}m_\mathrm{e}\cdot(y_\mathrm{e}/y^\mathrm{\textsc{sm}}_\mathrm{e}))^{-1/2}\gtrsim 110$~TeV.

Details on the status of ongoing FCC-ee monochromatization studies have been provided. The current monochromatization settings with largest Higgs signal strength correspond to $(\delta_{\sqrts},\LumiInt)\approx (7\,\mathrm{MeV},2\,\mathrm{ab}^{-1})$, and translate into a $0.4\sigma$ significance on the Higgs boson cross section, or correspondingly a $|y_\mathrm{e}|<2.6|y^\mathrm{\textsc{sm}}_\mathrm{e}|$ (95\% CL) upper bound, per IP and per year. 
Forthcoming extension and consolidation of this work, in the context of the anticipated FCC feasibility study, require at least the following activities:
\begin{description}
\item (i) Confirming the current signal significances with alternative MC event generators for the Higgs diboson backgrounds, in particular for the promising $\mathrm{H}\to\mathrm{W}\mathrm{W}^*\to\ell\nu 2j$ channel.
\item (ii) Studying the improvements of the FCC-ee detectors design needed in order to achieve the required accuracy and precision in key aspects of the analysis, such as the small light-quark-for-gluon mistagging efficiency of 1\% assumed in the key $\mathrm{H}\to gg$ channel.
\item (iii) Redoing the analysis using a more realistic (parametrized or full simulation) description of the detector response to more accurately assess the impact on the final signal significances of the reconstruction and selection efficiencies expected at FCC-ee.
\item (iv) Continuing and extending the accelerator monochromatization studies to improve the currently best FCC-ee working point of $(\delta_{\sqrts},\LumiInt)\approx(7\,\mathrm{MeV},2\,\mathrm{ab}^{-1})$, aiming at further reducing $\delta_{\sqrts}$ while increasing $\mathcal{L}$, and developing the corresponding optical lattices for the required beam optics parameters at the IP.
\end{description}

It is worth noting that running FCC-ee at $\sqrts = m_\mathrm{H}$ for a couple (or more) years can provide many more scientific outputs than the direct $s$-channel measurement considered here. Indeed, integrating tens of ab$^{-1}$ in $\epem$ collisions at 125\,GeV provides useful means to accurately determine the number of light neutrino families via $\mathrm{Z}(\nu\nu)\gamma$ radiative return~\cite{Gaemers:1978fe}, search for weakly-coupled BSM physics between the Z and Higgs mass poles~\cite{Agrawal:2021dbo}, and carry out other luminosity-demanding SM studies not accessible at the Z pole. 

In summary, the results presented in this essay demonstrate that FCC-ee is the most well-suited (if not, arguably, the unique) collider that can aim at a measurement of the electron Yukawa coupling via direct $s$-channel Higgs boson production. Such a measurement has many fundamental physics motivations and implications, among which: (i) it will explore the (so far hypothetical) Higgs mass generation mechanism for elementary particles of the first family of fermions that form the stable matter of the visible universe, (ii) it will scrutinize the electron's Yukawa coupling that, through its impact on the electron mass, sets the size of atoms and their energy levels (the Bohr radius is proportional to $1/m_\mathrm{e}$), (iii) it can access BSM scalar physics connected to the electron above the $\sim$100\,TeV scale, and (iv) it can directly probe the potential presence of any new particle that is quasi-degenerate (at the MeV level) with the Higgs boson mass.

\paragraph{Acknowledgements.}
We thank Roy Aleksan, Alain Blondel, Patrick Janot, and Frank Zimmermann for valuable discussions and feedback on various aspects of this study.

\bibliographystyle{myutphys}
\bibliography{FCCee_eeHiggs_denterria_2021}
\end{document}